\documentclass[aps,prb,twocolumn]{revtex4}

\usepackage{color}
\usepackage{amsmath}
\usepackage{amsthm}
\usepackage{amssymb}
\usepackage{graphicx}
\usepackage{tabularx}
\usepackage[squaren]{SIunits}

\let\a=\alpha  \let\g=\gamma  \let\e=\epsilon
   
\let\l=\lambda     
\let\s=\sigma   

\let\o=\omega

\def\be{\begin{equation}}
\def\ee{\end{equation}}
\def\bea{\begin{eqnarray}}
\def\eea{\end{eqnarray}}
\def\ba{\begin{array}}
\def\ea{\end{array}}

\def\ve{\varepsilon}

\newcommand{\bk}{{\bf k}}
\newcommand{\bko}{{\bf k}_1}
\newcommand{\bD}{{\bf D}}

\newcommand{\bp}{{\bf p}}
\newcommand{\bq}{{\bf q}}

\newcommand{\vko}{{\bf k_1}}
\newcommand{\vkt}{{\bf {k_2}}}

\newcommand{\bE}{{{\bf E}}}

\newcommand{\intk}{\int \frac{d^2\bk}{(2\pi)^2}}
\newcommand{\intko}{\int \frac{d^2\bk_1}{(2\pi)^2}}
\newcommand{\intq}{\int \frac{d^2\bq}{(2\pi)^2}}
\newcommand{\intp}{\int \frac{d^2\bp}{(2\pi)^2}}

\begin{document}

\title{Interaction dominated transport and Coulomb drag in bilayer graphene}

\author{Jonathan Lux}
\author{Lars Fritz}
\affiliation{Institut f\"ur Theoretische Physik, Universit\"at zu K\"oln, Z\"ulpicher Stra\ss e 77, 50937 K\"oln, Germany}

\date{\today}

\begin{abstract}
We investigate interaction effects in transport phenomena in bilayer graphene (BLG). For the minimal conductivity in pristine BLG, we find that the conductivity assumes
a constant value in the limit $T\to 0$, with the first correction being $\propto \sqrt{T}$. This has to be contrasted from the standard $1/T^2$ in Fermi liquids (neglecting additional logarithms and above all disorder).
We furthermore study the Coulomb drag resistivity between two BLGs in the whole range from deep within the Fermi liquid regime all the way to the charge neutrality (CN) point.
We find that in the Fermi liquid regime drag behaves very similarly to drag in a standard two-dimensional electron gas. In contrast to monolayer graphene, we find no saturation of drag as a function of the distance $d$ for realistic parameters.
In the vicinity of CN, we find an interesting interplay between interaction effects and disorder, like in the case of monolayer graphene. Here the drag resistivity strongly depends upon the ratio of the corresponding scattering times.
\end{abstract}


\maketitle


\section{Introduction}

It is well know that decreasing dimensionality is a means of increasing the effect of interactions in electronic systems.
In one dimension this leads to Luttinger liquid behaviour and a variety of manybody instabilities.\cite{luttinger} 
In two dimensions there also is a huge variety of strongly interacting electronic systems,\cite{twodim} leading to phenomena as diverse as the fractional quantum Hall effect \cite{FQHE1, FQHE2}
or high temperature superconductivity \cite{HTSC1, HTSC2}, to name two of the more prominent ones. 
A purely two-dimensional system by construction is monolayer graphene (MLG),\cite{Novoselov2004, Novoselov2005a} which is a hexagonal arrangement of carbon atoms. 
This system has remarkable electronic properties,\cite{RMPNeto} and a lot of these can be traced back to the fact that the low-energy theory assumes the form of an emergent massless Dirac equation.\cite{Novoselov2005b} 
In this system, while with magnetic field the fractional quantum Hall effect has been observed,\cite{Bolotin2009} without magnetic field the observation of interaction effects is scarce to date.
One of the reasons of the robustness of the underlying electronic system against interactions is the vanishing density of states at the Fermi level.\cite{Kotov2012} 

In bilayer graphene (BLG) systems the situation is different.\cite{mccann2012} The major difference stems from the fact that instead of massless Dirac particles, we deal with massive quasi--particles. \cite{mccann2006}
These occur since in the low-energy effective theory we have two bands which touch at one point, however not in a linear but in a quadratic fashion, implying a finite band mass.\cite{Novoselov2006}
As a consequence, even at charge neutrality (CN), the density of states is finite instead of zero as in MLG.
This implies that the system is much more unstable with respect to interactions. A variety of different symmetry broken states with a finite gap are conceivable,\cite{mccann2012}, for instance by explicitly breaking the interlayer symmetry by an external gate voltage.\cite{Ohta2006, Castro2007}
Without external perturbations there are indications that below $T=5$\,K BLG undergoes a spontaneous transition to a gapped state.\cite{Bao2012}
The exact nature of this state is currently under dispute \cite{Weitz2011, Freitag2012, Vafek2010, Zhang2010} and further theoretical and experimental activities are required to identify it unambiguously. 

Within this paper we do not attempt to speculate about the nature of this symmetry broken state but instead focus on interaction effects and their signature in transport properties.
Consequently our theory only applies for temperatures which are above the gap scale. It also neglects the scattering of electronic degrees from collective modes which can be precursors to the transition to the ordered state.  

In a first part, we investigate the minimal conductivity of clean BLG. As in the case of MLG there is a well-defined minimal conductivity, even in absence of impurities solely due to (screened) Coulomb interaction.
This is a consequence of the particle-hole symmetry in the model and holds true for every particle-hole symmetric system.
We study the minimal conductivity as a function of temperature, and find that it follows $\sigma_{\rm{min}}=\frac{e^2}{h} \left( 27.4 + 34.0  \frac{\ve}{e^2} \sqrt{\frac{T}{m}} \right)$ for all reasonable temperatures $T$.
In this expression $m$ is the effective mass of the BLG quasi--paticles, and $\epsilon$ is the dielectric constant of the environment.
Interestingly, it saturates for low temperatures as opposed to the standard behavior in Fermi liquids where it diverges (due to the vanishing phase space).  

In a second part we study the effect of Coulomb drag. This is a very direct measurement of Coulomb interaction between two electrically isolated two--dimensional electronic systems.
The experimental setup goes as follows: A current is driven in one of the layers, called the active one. Then Coulomb interaction can transfer momentum to the other layer, called the passive one.
If the current is not admitted to flow there, a voltage drop will be induced. This can be measured, and the ratio of the current in the active layer and the voltage drop in the passive layer is called the drag resistance.

Our study provides the first comprehensive study of Coulomb drag in BLG in all limits ranging from CN to the Fermi liquid (FL) regime.
To the best of our knowledge experimental results for such a setup are currently unavailable, and we are only aware of one theoretical study.\cite{hwang2011a}
For high doping we find that the system behaves like a standard FL, and we discuss all the limiting cases as a function of temperature $ T $ and distance $ d $ between the layers.
Unlike in the case of MLG, where there is a well-defined regime of 'zero' distance, this regime is lacking here due to the small screening length.
In the opposite limit, close to CN, we find that drag depends very sensitively on the ratio between relaxation due to interaction and disorder, thus inelastic and elastic scattering.
A similar behavior has recently been observed in theoretical studies of drag in parallel MLGs.\cite{we, Schuett2012}  

Technically, this paper follows in large part the lines of a recent publication \cite{we} on drag in a MLG setup and consequently we present little technical detail.
The generalization to the present situation is pointed out in the paper and straightforward if not explained otherwise.

\section{Model \& Method} \label{sec:model}

\subsection{The model}
The model Hamiltonian in our problem in general consists of two copies of the free BLG Hamiltonian for the active and passive layer, respectively, as well as interactions within and in-between layers. It reads
\begin{eqnarray}
H = \sum_{i=a,p} \left( H_0^i + H_{\rm{int}}^i + H_{\rm{dis}}^i  \right)+H^{ap}_{\rm{int}},
\end{eqnarray}
where $a$ denotes the active layer and $p$ the passive. $H_0^{a/p}$ denotes the free Hamiltonian in both layers, $H_{\rm{int}}^{a/p}$ the interaction within each layer, while $H^{ap}_{\rm{int}}$ describes the interaction between layers. Disorder is implemented within each layer via $H_{\rm{int}}^i$.

We restrict ourselves to a standard effective two-band description of BLG \cite{mccann2012}.
Around the points $K$ and $K'$ we can expand the dispersion and end up with the following effective Hamiltonian:
\begin{eqnarray} \label{eq:hk}
 H_0^i &=& \sum_{f=1}^N\int d^2 {\mathbf{k}}   \Psi_f^{i\dagger} \mathcal{H}({\mathbf{k}})  \Psi^{i\phantom{\dagger}}_f \quad{\rm{, with}} \nonumber \\ 
\mathcal{H}({\mathbf{k}})&=&\left ( \begin{array} {cc}  -\mu & \frac{1}{2 m} (k_x - i k_y)^2 \\ \frac{1}{2 m} (k_x + i k_y)^2 & -\mu  \end{array} \right) \; .
\end{eqnarray}
$ m = 0.054 \, m_e $ is the effective mass of BLG,\cite{Kotov2012} and $ m_e $ is the electron mass.
The two bands, subsequently identified with electron ($+$) and hole ($-$) band, have the respective energies, $ \e_k^\pm - \mu = \pm \frac{k^2}{2 m } - \mu $.
In this approximation, the system is particle--hole symmetric at CN ($ \mu = 0 $).
$ f $ sums over the spin and valley degrees of freedom, consequently in BLG we have $ N=4 $.
In the follwing, we maintain the parameter $ N $ as often as possible, to distinguish it from other numerical prefactors. In the numerical results it is always $ N = 4 $.
$ \Psi_{f}^{i [\dagger]} $ denotes the annihilation [creation] operator of layer $ i \in \{ a,p \} $ and flavor $ f \in \{ 1, \cdots, N  \} $.

It proves convenient for the following discussions to transform the Hamiltonian into the quasi-particle basis, where it reads
\begin{align} \label{eq:h0}
 H_0^i = \sum_{\l = \pm} \sum_{f=1}^N \intk \frac{\l \, k^2}{2 m}\;  \g_{\lambda, f}^{i\dagger } (\bk) \g_{\l, f}^{i\phantom{\dagger}} (\bk)\;.
\end{align}
$ \g_{\l ,f}^{i [\dagger]} $ denotes the annihilation [creation] operator of $ \l \in \{ +, -\}$ quasi--particles in layer $ i \in \{ a,p \} $ with flavor $ f \in \{ 1, \cdots, N  \} $.

Since we are interested in conductivities we express the current in the particle--hole basis. As in the case of MLG the current decomposes into two contributions
\begin{eqnarray}
{\bf{J}}={\bf{J}}_{\rm{I}}+{\bf{J}}_{{\rm{II}}}
\end{eqnarray}
where the first accounts for the motion of the quasi--particles, while the second is the incoherent contribution. In terms of the electron and hole operators they read
\begin{eqnarray}
{\bf{J}}_{\rm{I}}&=&\frac{e}{m}\sum_{\lambda f}\int \frac{d^{2}k}{(2\pi )^{2}}
\lambda \mathbf{k}\gamma _{\lambda f}^{\dagger }(\mathbf{k})\gamma
_{\lambda f}(\mathbf{k})\,,   \nonumber \\ {\bf{J}}_{\rm{II}}&=&-i\frac{e}{m}\sum_f\int \frac{d^{2}k}{(2\pi )^{2}}(\hat{\mathbf{z}%
}\times \mathbf{k}) \nonumber \\ &\times& \left[ \gamma _{+f}^{\dagger }(\mathbf{k})\gamma
_{-f}(\mathbf{k})-\gamma _{-f}^{\dagger }(\mathbf{k})\gamma _{+f}(\mathbf{k}%
)\right]\;,
\end{eqnarray}
where ${\mathbf{z}}$ is the unit vector perpendicular to the bilayer plane.
Like in the case of MLG the d.c. conductivity is dominated by ${\bf{J}}_{\rm{I}}$ corresponding to the quasi--particle contribution which produces a Drude peak in the absence of disorder and interactions.
In the following we investigate the broadening of this quasi--particle peak.

\subsubsection{Coulomb interaction, screening, and disorder}

We consider the effect of Coulomb interaction, which in two dimensions is given by: 
\begin{equation} \label{eq:V1}
 V(q) = \frac{2 \pi e^2}{\ve q} \; ,
\end{equation}
where $ \ve $ is the dielectric constant of the surrounding medium.
The Coulomb interaction between the two layers separated by the distance $ d $, which mediates the Coulomb drag, reads:
\begin{equation} \label{eq:U1}
 U(q) = \frac{2 \pi e^2}{\ve q} e^{- q d} \; .
\end{equation}

Unlike MLG undoped BLG has a finite density of states at the CN point. Like in the two--dimensional electron gas (2DEG), this is constant, and given by
\begin{equation} \label{eq:dos}
 \nu (E) = N \intk \delta \left( \frac{k^2}{2 m} - E \right) = N \frac{m}{2 \pi} \; .
\end{equation}
This implies that Coulomb interactions are always screened and in the static limit $ \Pi (q, \o = 0) = \nu $ the Coulomb interaction within the random-phase approximation (RPA) becomes
\begin{align} \label{eq:V1rpa}
 V_{RPA} (q) = \frac{V}{1+V \Pi} = \frac{2 \pi e^2}{\ve (q+N q_{TF})} \; ,
\end{align}
where we have defined the inverse Thomas-Fermi screening length
\begin{equation} \label{eq:qtf}
 q_{TF} = m e^2/\ve \approx \frac{\ve_0}{\ve} 10^{10} \; \text{m}^{-1} \; .
\end{equation}
This corresponds to very small distances and defines the Bohr radius, which becomes important subsequently.

The RPA for the two--layer setting is more involved and reads for the intra--layer interaction
\begin{align} \label{eq:V2rpa} 
 V^{a / p}_{2RPA} & = \frac{V+ (V^2 - U^2)\Pi_{p / a}}{\left(1+V \Pi_a\right)\left(1+V \Pi_p\right)-U^2 \Pi_a \Pi_p} \\ 
  & = \frac{2 \pi e^2 }{\ve} \frac{q + (1-e^{- 2 q d}) N q_{TF} }{(q+N q_{TF})^2 - e^{-2 q d} N^2 q_{TF}^2} \nonumber \; .
\end{align}
For the inter--layer interaction we find
\begin{align} \label{eq:U2rpa} 
  U_{2RPA} & = \frac{U}{\left(1+V \Pi_a\right)\left(1+V \Pi_p\right)-U^2 \Pi_a \Pi_p} \\
  & = \frac{2 \pi e^2 }{\ve} \frac{q e^{- q d}}{(q+ N q_{TF})^2 - e^{-2 q d} N^2 q_{TF}^2} \nonumber \; .
\end{align}

In order to express the Coulomb interactions in the quasi--particle basis it proves useful to introduce the coherence factor
\begin{equation} \label{eq:vertex}
M_{\l , \l^\prime} (\bk, \bq) = \frac{1}{2} \left( 1 + \l \l^\prime \frac{Q \cdot K^\ast}{Q^\ast \cdot K} \right) \; ,
\end{equation}
where $ K = k_x + i ky $ and $ Q = q_x + i q_y $. Together with 
\begin{align} \label{eq:Tintra}
 & T_{\l_1, \l_2, \l_3, \l_4 } (\bk, \bp, \bq)   \nonumber \\
= & \frac{V(q)}{2} M_{\l_1 , \l_4} (\bk, \bk+\bq) M_{\l_2 , \l_3} (\bp, \bp-\bq) \; 
\end{align}
this allows to express the intra-layer interaction in the quasi--particle basis
\begin{widetext}
\begin{align} \label{eq:hintra}
H_{int}^{aa} = \sum_{\l_i, f, f^\prime} \intk \intp \intq T_{\l_1, \l_2, \l_3, \l_4 } (\bk, \bp, \bq) \; \g^\dagger_{a,\l_4, f} (\bk+\bq) \g^\dagger_{a,\l_3, f^\prime} (\bp-\bq) \g_{a,\l_2, f^\prime} (\bp) \g_{a,\l_1, f} (\bk) \; .
\end{align}
For the inter-layer interaction we define
\begin{align} \label{eq:Tinter}
  \widetilde{T}_{\l_1, \l_2, \l_3, \l_4} (\bk, \bp, \bq)  =  \frac{U(q)}{2} M_{\l_1 , \l_4} (\bk, \bk+\bq) M_{\l_2 , \l_3} (\bp, \bp-\bq) \; ,
\end{align}
where $ U $ depends on the distance $ d $ betweeen the layers. Then 
\begin{align} \label{eq:hinter}
H_{int}^{ap} = \sum_{\l_i, f, f^\prime} \intk \intp \intq \widetilde{T}_{\l_1, \l_2, \l_3, \l_4 } (\bk, \bp, \bq) \; \g^\dagger_{a,\l_4, f} (\bk+\bq) \g^\dagger_{p,\l_3, f^\prime} (\bp-\bq) \g_{p,\l_2, f^\prime} (\bp) \g_{a,\l_1, f} (\bk) \; .
\end{align}
\end{widetext}
If the system is at CN, there is no need to introduce impurities in order to have well defined conductivities since the current carrying state is at zero total momentum and thus no momentum relaxation is needed.
However, for finite chemical potential the current carrying states excites a global momentum and we have to take into account breaking of translational invariance due to the presence of impurities to relax the momentum. We model disorder as charged impurity scattering with the Hamiltonian
\begin{widetext}
\begin{align} \label{eq:himp}
  H_{imp} = \sum_i \sum_{\l, \l^\prime = \pm} \sum_{f=1}^N \intk \intq \frac{2 \pi e^2}{\e (|\bk - \bq|+N q_{TF})} M_{\l , \l^\prime} (\bk, \bq)  \g_{\lambda, f}^{\dagger } (\bk) \g_{\l^\prime, f} (\bq) \exp[i{\bf x}_i\cdot(\vko-\vkt)] \; .
\end{align}
\end{widetext}

\section{Kinetic approach}
We use the variational principle to calculate conductivities from linearized Boltzmann equation as described in Chapter 7 of Ref.~\onlinecite{zimanbook}. Generalizations to drag are straightforward and can for instance be read in Ref.~\onlinecite{we}.

The one--particle distribution functions of electrons ($+$) and holes ($-$) in the layer $i \in \{ a, p \} $ is defined as
\begin{equation}
f^i_{\lambda }(\mathbf{k},t)=\left\langle \gamma_{\lambda f}^{i\dagger }(%
\mathbf{k},t)\gamma_{\lambda f}^{i\phantom{\dagger}}(\mathbf{k},t)\right\rangle .  \label{defg}
\end{equation}%
We use the following ansatz for the non--equlibrium distribution function
\begin{equation} \label{eq:ansatz}
 f_\l = f^0_\l + \chi_\l \;  f^0_\l (1-f^0_\l) \frac{1}{m T^2} \bk \cdot e \bE \; ,
\end{equation}
where $ f_\l^0 $ is the equilibrium distribution function (Fermi function)
\begin{equation} \label{eq:fermifunc}
 f_\l^0 (\bk) = \frac{1}{1+ e^{(\e_k^\l - \mu)/T }} \; .
\end{equation}
We will use a one mode approximation, \eqref{eq:ansatz}, taking into account only the momentum mode $ \sim k $.
One can easily incorporate other modes like $ \sim k^0, k^2, k^3 $ into the formalism.
We have checked in the limiting cases of our study, that including these modes gives modified numerical values with corrections smaller than 1 per cent.
Consequently, we have discarded them throughout. 

The collision integrals, which are part of the Boltzmann treatment, are shown explicitly in Appendix~\ref{app:collision}.
They can be derived directly from the interaction terms via Fermi's Golden rule or equivalently from a Keldysh calculation.\cite{Fritz2008}

\subsection{Drag resistivity}
We calculate the drag resistivity from the conductivities obtained from a Boltzmann approach, see App.~\ref{app:collision}.
We consider the response tensor which has a structure similar to the one in the Hall effect.
The electric field ${\bf{E}}_a$ is only applied in the active layer $a$, and induces a current in the active layer called ${\bf{j}}_a$. 
It can also induce a current ${\bf{j}}_p$  in the passive layer, which is the drag effect.
Consequently, there are layer--diagonal and layer off--diagonal conductivities involved:
\begin{eqnarray}
\left ( \begin{array} c {\bf{j}}_a \\  {\bf{j}}_p \end{array} \right) = \left (  \begin{array} {cc}  \sigma_a & \sigma_d \\  \sigma_d & \sigma_{p}   \end{array}  \right) \cdot \left ( \begin{array} c {\bf{E}}_a \\  {\bf{0}} \end{array} \right) \;.
\end{eqnarray}
This tensor includes the individual conductivities $\sigma_a$ and $\sigma_p$ of the active and of the passive layer, respectively. The layer off--diagonal part is called the drag conductivity, denoted by $\sigma_d$, and it corresponds to a cross--conductance.
In the concrete experiment, however, the boundary conditions are such that the passive layer does not carry current.
Demanding ${\bf{j}}_p={\bf{0}}$ requires a field counteracting the flow in the passive layer which is given by ${\bf{E}}_p=-\frac{\sigma_d}{\sigma_p}{\bf{E}}_a$. This implies that the drag resistivity (or transresisitivity) is given by
\begin{eqnarray}\label{eq:dragres}
\rho_d=\frac{|{\bf{E}}_p|^2}{{\bf{j}}_a \cdot {\bf{E}}_p}=\frac{-\sigma_d}{\sigma_a \sigma_p-\sigma_d^2}\;.
\end{eqnarray}

Like in the case of thermal transport $\rho_d$ can be finite even if the individual conductivities $\sigma_a$, $\sigma_p$, and $\sigma_d$ diverge.\cite{we}
This is an effect of the boundary condition of vanishing charge flow in the passive layer analogous to a finite thermal conductivity in thermal transport in Fermi liquids.\cite{thermal}

\section{Single layer conductivity}
In this section we describe the conductivity of an isolated BLG sheet. In the drag setup this limit is naturally achieved by setting the distance $d$ between active and passive layer to infinity.
As in the case of MLG this quantity is well defined at the CN point, even in absence of impurities. This is a direct consequence of the particle-hole symmetry in the system, which
implies that the current carrying state has effectively zero total momentum.

The interaction limited single--layer conductivity has been investigated in MLG with the result \cite{Fritz2008} $\s = 0.76 \; \frac{e^2}{h \, \a(T)^2}$.
In this expression $\alpha=\frac{e^2}{\epsilon v_F}$ corresponds to the fine structure constant of graphene. It has been shown \cite{RGMLG1, RGMLG2} that it has a logarithmic flow to zero upon decreasing temperature $T$.
In the large N limit it was found \cite{Kashuba2008} that the numerical prefactor is given by $0.69$. While in MLG at CN it is justified to neglect the effect of screening due to the lack
of density of states, this is not true for BLG. To see this one has to compare the temperature $T$ against the energy associated with the Thomas-Fermi wavevector, which is given by
\begin{equation}
 \frac{q_{TF}^2}{m} \approx \left( \frac{\ve_0}{\ve} \right)^2 \times 10^4 \;\text{K} \; .
\end{equation}
Thus for realistic temperatures one is always in the limit $ T m \ll q_{TF}^2 $. Consequently, one always has to consider screening and we use the screened version
of Coulomb interaction in the collision integral of the Boltzmann equation. 

Solving the Boltzmann equation numerically in the one mode approximation yields the following expression for the minimal conductivity :

\begin{figure}[ht]
\includegraphics[width=0.45\textwidth]{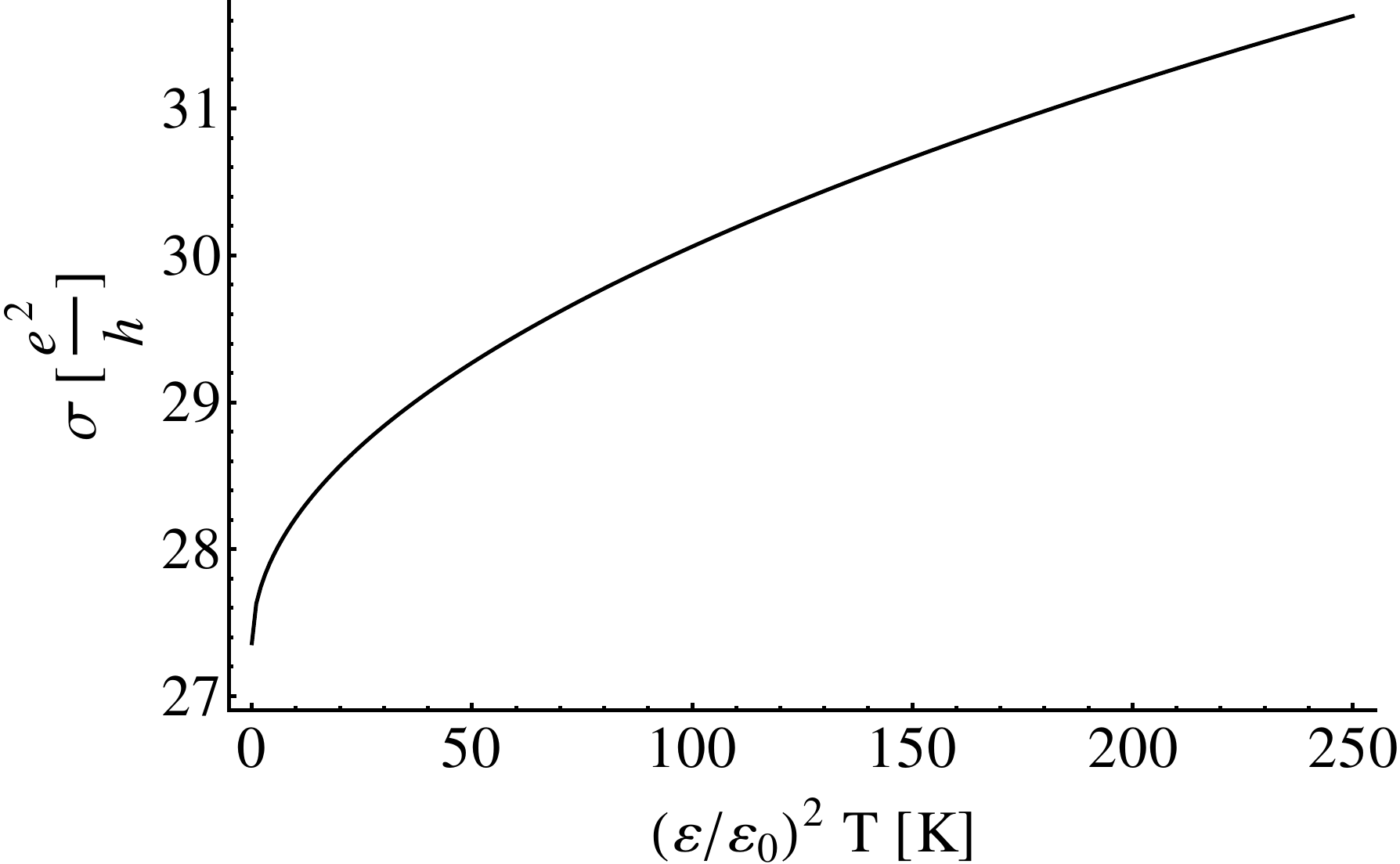}
\caption{Interaction dominated single-layer conductivity. The leading temperature behavior is of the unusual $\sqrt{T}$ type. In the limit $T \to 0$, $ \s $ converges to a fintie value $ \sigma (T \to 0) = 27.4 \, \frac{e^2}{h} $.
The limiting value cannot be reached in experiment as a band gap opens below $ T \approx 5 $\, K and our description breaks down.\cite{Bao2012}}\label{fig:s0_tdep}
\end{figure}
\begin{align} \label{eq:s0BLG_Tsmall}
 \s & \approx \frac{e^2}{h} \left( 27.4 + 34.0  \frac{\ve}{e^2} \sqrt{\frac{T}{m}} \right) \nonumber \\
    & = \frac{e^2}{h} \left( 27.4 + 0.259  \frac{\ve}{\ve_0} \sqrt{T \, [ \text{K} ]} \right) \; . \nonumber 
\end{align}
This implies that there is a constant conductivity in the limit of zero temperature and the leading temperature behavior is an unusual $\sqrt{T}$. This can be rationalized as follows: assuming $ q_{TF}^2 \gg T m $ the typical structure of integrals in the collision integral is given by.
\begin{equation} \label{eq:sqrtint}
\left( \int dq \, \frac{e^{- q^2}}{(\sqrt{T m} q + q_{TF})^2} \right)^{-1} \sim q_{TF}^2 + \sqrt{T m} \, q_{TF} \; ,
\end{equation} where the exponential factor accounts for the Fermi distribution function.
The screened Coulomb interaction acts like a local interaction and dimensional analysis leads to a minimal conductivity which is independent of temperature, in agreement with the full numerical solution.
The leading correction is linear in $q$ and consequently consequently produces a behavior corresponding to $\propto \sqrt{T}$ (note that one can again understand this from dimensional analysis taking into account that the dynamical exponent is $z=2$).
There also is a high temperature limit, in which the conductivity behaves linear in $T$. While this is a theoretical possibility this behavior dominates for temperatures above $10^4$ K, and consequently plays no role in experiments. Also, the zero temperature limit cannot be reached in experiment since below $T=5$\;K BLG supposedly is gapped and our analysis breaks down.
The result of the full numerical calculation of the conductivity is shown in Fig.~\ref{fig:s0_tdep}, and the agreement with the $\sqrt{T}$ behavior is excellent within the relevant temperature range.

\section{Coulomb drag}
Motivated by recent experiments \cite{Kim2011, Kim2012, Gorbachev2012} many theoretical studies of Coulomb drag
in MLG have been published.\cite{we, hwang2011a, Tse2007, Narozhny2007, Katsnelson2011, Peres2011, Narozhny2011, Polini2012, Schuett2012, Levitov2012}
We now investigate Coulomb drag in BLG and compare the results to those found for MLG.

\subsection{Qualitative results}

Our subsequent discussion has a number of natural dimensionless parameters associated with different regimes. First and foremost, the parameter $\mu/T$ allows to distinguish whether we are in the FL regime ($\mu/T \gg1 $)
or in the limit close to CN ($\mu/T \ll1$). The parameter $\frac{q_{TF}^2}{mT}$ is always much larger than one for practical purposes,
and consequently will not be discussed explicitly. We always assume Coulomb interaction to be screened. A further parameter important for the screening of the inter--layer interaction is given by $d q_{TF}$.
Due to $q_{TF} \sim 10^{10} \, \text{m}^{-1}$, which is the inverse Bohr radius, we are always in the limit $d q_{\rm{TF}} \gg 1 $. The parameter which allows to distinguish a clean system, meaning that the single--layer conductivities are dominated by 
Coulomb scattering, from a disordered system, where the single--layer conductivities are set by impurity scattering, is given by $g=\frac{\hbar^2 n_{\rm{imp}}}{Tm}$. This corresponds to ratio of the impurity concentration, $n_{\rm{imp}}$, to
the typical thermal momentum squared, or, equivalently, to the ratio of the scattering times associated with impurity scattering and Coulomb scattering.

\begin{figure}[h]
\includegraphics[width=0.45\textwidth]{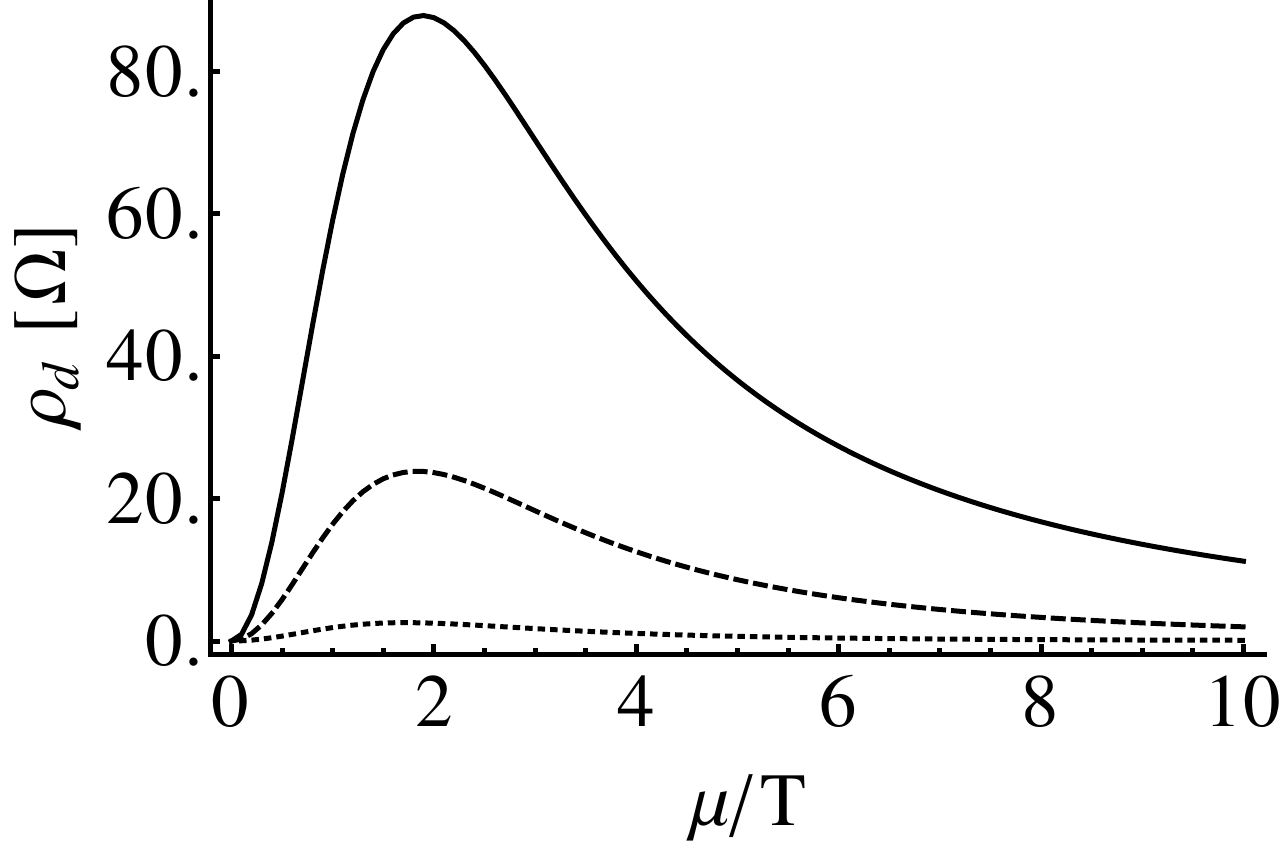}
\caption{Drag resistivity for $ \frac{q^2_{TF}}{ m T} = 100 $ and $ d q_{TF} = 1 $ (solid line), $ d q_{TF} =2 $ (dashed line) , $ d q_{TF} = 5 $ (dotted line).
The single--layer conductivities $ \s_a, \s_p $ are disorder dominated ($ g = 1000 $). Then the maximum is located at $ \mu \sim 2 T $.}
\label{fig:drag_qtf10}
\end{figure}

In order to get the overall qualitative picture we have plotted the full crossover curve of drag in Fig.~\ref{fig:drag_qtf10}. We observe that it is very reminiscent of the full crossover curve in MLG.
However, as we will point out, there are important differences, mainly due to the fact that even in pristine BLG there is an additional energy scale due to the finite band mass.
This mostly shows up in the distance dependence in the FL regime as well as on the behavior upon approaching the CN point. 

\subsection{FL regime: $\mu/T \gg 1$} \label{subsec:dragFL}

In the case of drag in MLG the effect of impurity scattering cancels exactly from $\rho_d $ in the FL regime. 
Consequently, all information about disorder is gone (this equally applies to a 2DEG). Using the same steps and manipulations as in the MLG,\cite{we} one can show that within the one-band approximation
\begin{widetext}
 \begin{align} \label{eq:rhodFL}
   \rho_d(\mu/T \gg1 )= \frac{\hbar}{e^2} \frac{4}{\pi}  \frac{1}{T \mu_a \mu_p}  \frac{1}{\nu_a \nu_p} \int d\omega \int \frac{d^2q}{(2\pi)^2}  \frac{q^2}{\sinh^2 \frac{\omega}{2T}} |U_{ap}({\bf{q}},\omega))|^2  
\operatorname{Im} \Pi^{++}_{p} ({\bf{q}},\omega) \operatorname{Im} \Pi^{++}_{a} (-{\bf{q}},-\omega)  \; ,
 \end{align}
\end{widetext}
where the density of states $ \nu_{a,p} $ was defined in Eq.~\eqref{eq:dos} and $ U_{ap} $ is the screened inter-layer interaction, Eq.~\eqref{eq:U2rpa}.
For $ |\mu| \gg T $ the polarization function $\Pi^{++}$ (the $++$ refers to the fact that only the quasi--particle contribution of the majority charge carrier is taken into account) assumes the standard Fermi-liquid form given by
\begin{equation} \label{eq:impi++}
 \operatorname{Im} \Pi^{++} (q, \o) \approx \nu \frac{m \o}{k_F q} \Theta \left( 2 k_F - q \right)  \Theta \left( \frac{k_F q}{m} - |\o| \right) \; .
\end{equation}

Scaling $ \o \to \o T $ and $ q \to q/d $ allows to identify three natural dimensionless parameters:
$ \overline{d} = T d m/ k_F $, $ d k_F $ and $ d q_{TF} $.
We can then write the drag resistivity as a prefactor times a function which only depends on dimensionless parameters, namely
\begin{equation} \label{eq:rhodFLscaled}
 \rho_d = \frac{h}{e^2} \frac{4}{\pi} \left( \frac{T}{\mu} \right)^2 \left( \frac{q_{TF}}{k_F} \right)^2 F(\overline{d}, d k_F, d q_{TF}) \; 
\end{equation}
with
\begin{widetext}
\begin{align} \label{eq:rhodFLg}
 F (\overline{d}, d k_F, d q_{TF}) & = \int\limits_0^{2 d k_F} d q \frac{q^3 e^{-2 q}}{((q+N d q_{TF})^2 - e^{- 2 q} (N d q_{TF})^2)^2}
 \;  \int\limits_{-q/\overline{d}}^{q/\overline{d}}  d \o  \frac{\o^2}{\operatorname{\sinh}^2 \left( \o/2 \right) } \; .
\end{align}
\end{widetext}
Since we have three independent parameters, we expect to find eight different regimes, depending on whether the parameters are much larger or smaller than unity. 
Two of these limits are internally incompatible with each other and there remain six regimes. Among these six only three are physical in the sense that for realistic systems they can be achieved.
For more details regarding the analysis of the integral we again refer the reader to Ref.~\onlinecite{we}.

\subsubsection{$\overline{d}\ll1$}

There are four limiting cases in this regime, two of them being physically sensible. We first discuss the situation of $k_F d \gg 1$. In that situation we have either $dq_{TF}\gg 1$ where the drag resistivity reads
\begin{equation} \label{eq:rhodFLlgg}
 \rho_d = \frac{h}{e^2} \frac{4 \pi \zeta(3)}{N^4} \left( \frac{T}{\mu} \right)^2 \frac{1}{(d q_{TF})^2} \frac{1}{(d k_F)^2}
\end{equation} 
or $dq_{TF}\ll 1$ where we find 
\begin{equation} \label{eq:rhodFLlgl}
 \rho_d = \frac{h}{e^2} \frac{32 \pi}{3} \left( \frac{T}{\mu} \right)^2  \left( \frac{q_{TF}}{k_F} \right)^2 \ln \left( \frac{1}{2 N d q_{TF}} \right)\;.
\end{equation}
If we go into the limit $ k_F d \ll 1 $ we find in the limit $ d q_{TF} \ll 1 $
\begin{align} \label{eq:rhodFLlll} 
 \rho_d & = \frac{h}{e^2} \frac{32 \pi}{3} \left( \frac{T}{\mu} \right)^2  \left( \frac{q_{TF}}{k_F} \right)^2 \times \nonumber \\ 
& \times \left( \ln \left( \frac{N q_{TF}+ k_F}{N q_{TF}} \right) - \frac{k_F}{N q_{TF} + k_F} \right)
\end{align}
while in the limit $ d q_{TF} \gg 1 $ there is
\begin{equation} \label{eq:rhodFLllg}
 \rho_d =  \frac{h}{e^2} \frac{16 \pi}{3 N^4} \left( \frac{T}{\mu} \right)^2 \frac{1}{(d q_{TF})^2} \;.
\end{equation}

While the limit $d q_{TF} \ll 1$ theoretically makes sense it is not to be achieved in experiments due to the small screening length $ \sim q_{TF}^{-1} $.
Consequently, there will be no saturation of distance dependence as in the case of MLG.\cite{we, Narozhny2011, Polini2012}

\subsubsection{$\overline{d}\gg 1$}

Here we find only two liming cases the reason being that this limit is incompatible with $dk_F \ll 1$. As in the case of MLG, this regime is characterized by a linear in $T$ behavior.
In the limit $k_F d \gg 1 $ we have two situations, one in which $ d q_{TF} \gg 1 $ with 
\begin{equation} \label{eq:rhodFLggg}
 \rho_d =  \frac{h}{e^2} \frac{8 \pi3}{15 N^4} \left( \frac{T}{\mu} \right) \frac{1}{(d q_{TF})^2} \frac{1}{(d k_{F})^3} 
\end{equation}
while for $ d q_{TF} \ll 1 $ we have
\begin{equation} \label{eq:rhodFLggl}
 \rho_d = \frac{h}{e^2} \frac{32}{\pi} \left( \frac{T}{\mu} \right) \left( \frac{q_{TF}}{k_F} \right)^2 \frac{1}{d k_{F}} \;.
\end{equation}

Again it is important to note that the last regime is only theoretically possible.

\subsubsection{Summary of results}

From the preceding discussion we conclude that in experiment only three different regimes can be realizable:
\begin{eqnarray}\label{eq:summary}
\rho_d \approx \left \{ \begin{array} {ccc}      \frac{h}{e^2} \frac{16 \pi}{3 N^4} \left( \frac{T}{\mu} \right)^2 \frac{1}{(d q_{TF})^2}  & \overline{d} \ll 1  &  k_F d \ll 1 \\  
 \frac{h}{e^2} \frac{4 \pi \zeta(3)}{N^4} \left( \frac{T}{\mu} \right)^2 \frac{1}{(d q_{TF})^2} \frac{1}{(d k_F)^2} &  \overline{d}\ll 1 & k_F d \gg 1 \\  
  \frac{h}{e^2} \frac{8 \pi3}{15 N^4} \left( \frac{T}{\mu} \right) \frac{1}{(d q_{TF})^2} \frac{1}{(d k_{F})^3}  & \overline{d} \gg 1 & k_F d \gg 1   \end{array}  \right. \;.\nonumber \\
\end{eqnarray}
Unlike MLG there is no regime in which drag becomes independent of the distance $d$.

\subsection{Close to charge neutrality: $\mu/T \ll1$}

In this limit drag resistivity is very sensitive to the ratio of the inelastic scattering time due to Coulomb interaction and the elastic impurity scattering time. This behavior has recently been analyzed for MLG \cite{we, Schuett2012} and qualitatively the result in the BLG case is similar.
The drag resistivity is well--defined and finite even in the clean system at finite chemical potential, where the individual layer-- and drag conductivities diverge.
The parameter which allows to go from the dirty limit to the clean limit is given by 
\begin{equation} \label{eq:gdef}
 g = \frac{\hbar^2 n_{imp}}{T m} \; .
\end{equation}
For $g\to 0$ we realize the clean limit, while for $g \to \infty$ we are in the dirty limit. In contrast to the FL regime, where elastic scattering drops out exactly, we find that the drag resistivity depends strongly on $g$.
For disorder dominated single--layer conductivities, $ g \gg 1 $, we find a maximum in the drag resistivity at $ \mu \sim 2 T $, see Fig.~\ref{fig:drag_qtf10}. The same was shown to hold in MLG.\cite{we, Narozhny2011, Schuett2012}
For smaller values of $ g $, the maximum shifts to smaller values of $ \mu/T $. In the following we use the parameter $\overline{\mu}_{a,p}=\mu_{a,p}/T$ for both layers, active and passive, individually as the dimensionless chemical potential.

It was pointed out in the context of MLG\cite{we,Schuett2012} that there are now two orders of limits in which one can describe Coulomb drag in a clean system at particle--hole symmetry.
If one first considers the system at CN, $ \mu=0 $, drag vanishes irrespective of the disorder strength by virtue of lack of momentum transfer.
However, if one first extrapolates to the clean system, $g\to 0$, at finite chemical potential and afterwards takes the limit $\mu \to 0$ one ends up with a finite drag resistivity.
Like in the case of MLG the first way to take the limits is the physical one, resulting in a vanishing drag at particle--hole symmetry. In total, this shows that the drag resistivity sensitively depends upon the ration between inelastic and elastic scattering in the vicinity of CN.

\begin{figure}[tbh]
\includegraphics[width=0.45\textwidth]{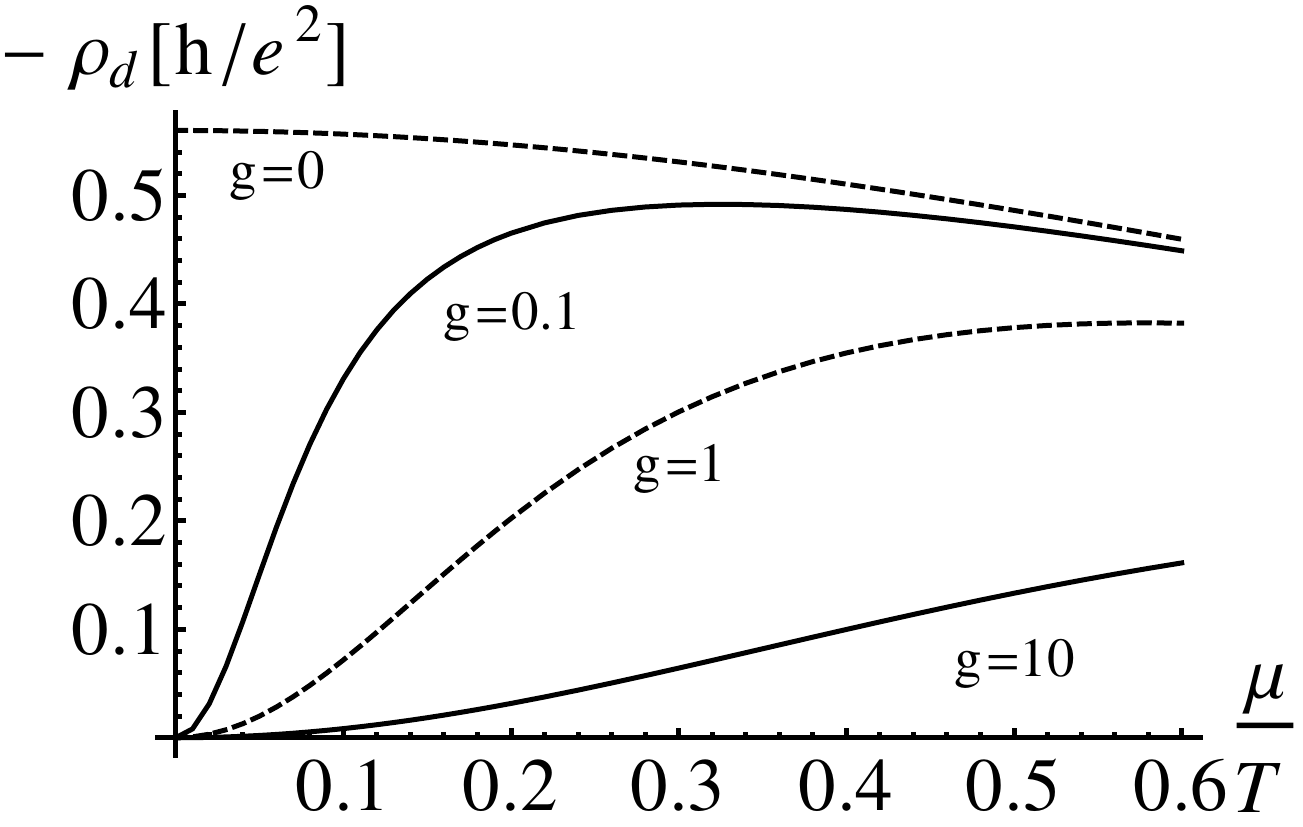}
\caption{Drag resistivity for a series of different values of $g$ in the idealized zero distance limit, $ d q_{TF}= 0 $ ($\frac{q_{TF}^2}{mT}=100$).
The single layer conductivities $ \s_a, \s_p $ range from interaction dominated $g \ll 1$ with a maximum at $\mu/T<2$ to disorder dominated for $g \gg 1$ in which case the maximum is located at $ \mu \approx 2 T $. For $g=0$ the maximum pushes to zero.
For $ \mu \gtrsim 2 T $ the curves become independent of $ g $ and collapse to a single one, see section \ref{subsec:dragFL}.}
\label{fig:drag_musmall}
\end{figure}

This behavior can best be exemplified in the unphysical zero distance limit, $d q_{TF}=0$, which we show in Fig.~\ref{fig:drag_musmall}.  

\begin{figure}[tbh]
\includegraphics[width=0.45\textwidth]{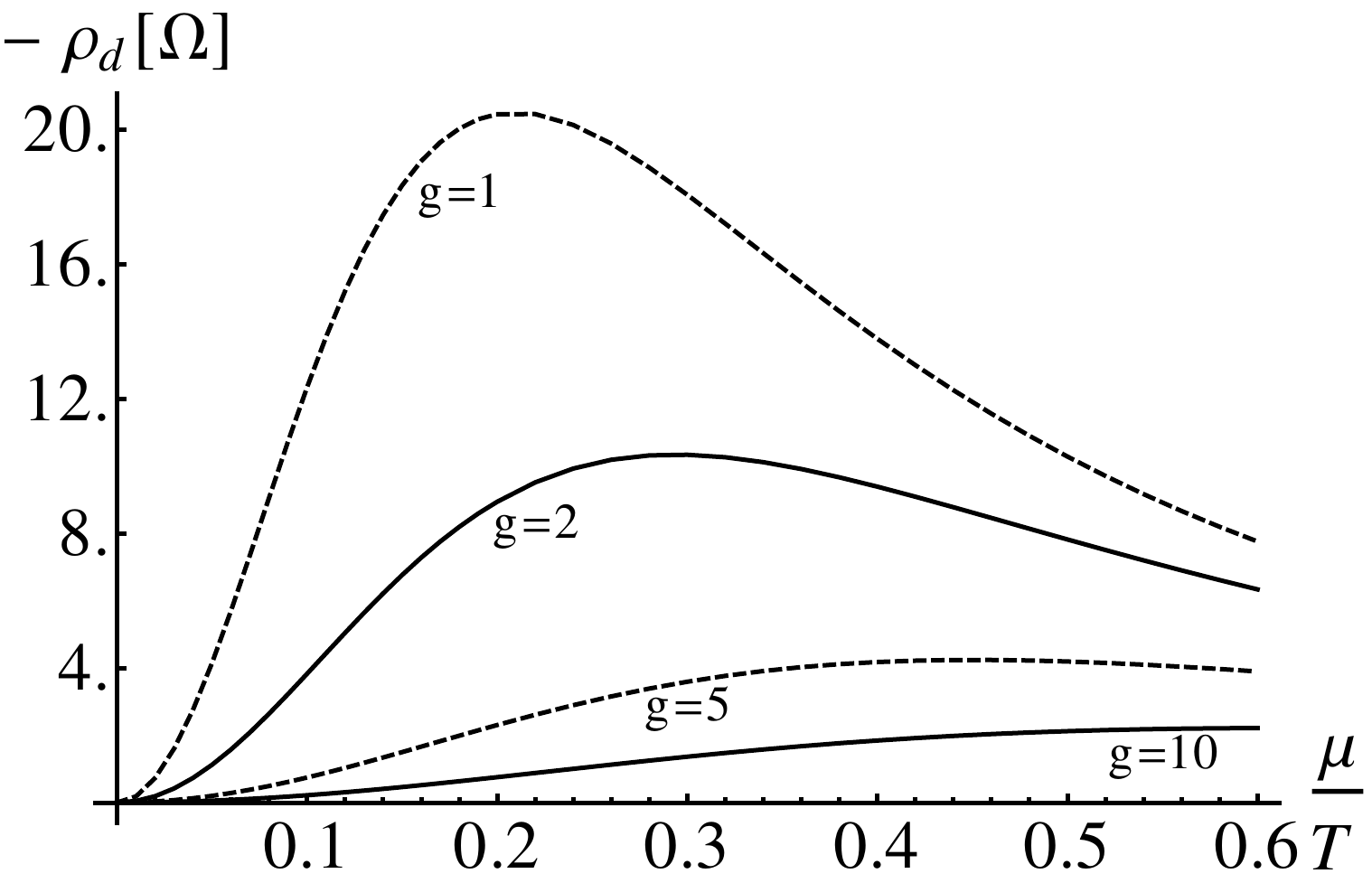}
\caption{Drag resistivity for $ \frac{q^2_{TF}}{ m T} = 100 $ and reasonable distances $ d q_{TF}= 10 $. The single layer conductivities $ \s_a, \s_p $ here are mosty interaction dominated. This can be seen from the fact that the maximum is located at $ \mu < 2 T $, in contast to the disorder dominated
case, Fig.~\ref{fig:drag_qtf10}.
For $ \mu \gtrsim 2 T $ the curves become independent of $ g $ and collapse to a single one, see section \ref{subsec:dragFL}.}
\label{fig:drag_mus_dfin}
\end{figure}

The general behavior assumes the following form: For small values of these parameters $\mu_a$ and $\mu_p$ in the individual layers, we find a behavior which is well described by the function
\begin{equation} \label{eq:rhod_musmall_fit}
 \rho_d \approx - \frac{h}{e^2} \frac{\overline{\mu}_a \overline{\mu}_p}{a_1 (g, d q_{TF}, \frac{q_{TF}^2}{T m}) + a_2 (g, d q_{TF}, \frac{q_{TF}^2}{T m}) \overline{\mu}_a \overline{\mu}_p} \; ,
\end{equation}
If we now keep $g$ finite and consider $ |\mu_{a,p}|/T \ll 1 $ we find
\begin{align} \label{eq:rhod_musmall_gfin}
\rho_d & = \frac{e^2}{h} \; \frac{\overline{\mu}_a \overline{\mu}_p}{a_1(g)} \nonumber \\
& = - \frac{e^2}{h} \; \frac{\mu_a \mu_p}{T^2} \; \frac{b_1 T m + b_2 n_{imp}}{n_{imp}} \; ,
\end{align}
where $ b_1 , b_2 $ depend on $ q_{TF}^2 /(Tm) $ and $ d q_{TF} $. This expression goes to zero if we extrapolate the chemical potentials to zero. 

On the other hand, in the clean limit $ g \to 0 $ we find $ a_1 \to 0 $. Now
\begin{align} \label{eq:rhod_musmall_g0}
\rho_d & = \frac{e^2}{h} \; \frac{\overline{\mu}_a \overline{\mu}_p}{a_2(g) \overline{\mu}_a \overline{\mu}_p} \nonumber \\
& = - \frac{e^2}{h} \; \frac{1}{a_2 (g)} \; \underset{g \to 0}{\to} \; c \; \frac{e^2}{h} \; ,
\end{align}
meaning it extrapolates to a finite value as $\mu_{a,p} \to 0$. 
The constant $ c \neq 0 $, in contrast to the MLG\cite{we, Schuett2012}, is not a universal number but it depends on $ q_{TF}^2 /(Tm) $ and $ d q_{TF} $, but not on $ g $ anymore (since we have performed the limit $g\to 0$ first).

We have furthermore plotted a sequence of crossover curves with decreasing disorder strength in Fig.~\ref{fig:drag_mus_dfin} for a more realistic value of $d q_{TF}=10$.
As one can see form these curves, it is possible to  extract information about both inelastic and elastic scattering times from Coulomb drag experiments carried out in the vicinity of CN. This has to be contrasted from the FL regime, in which information about disorder is absent.

\section{Conclusion}

In this paper we have investigated interaction effects in transport phenomena in BLG. In a first part we studied the minimal conductivity in pristine BLG. There we find that the conductivity assumes a constant value in the limit $T\to 0$. Importantly, this result is valid in the clean limit and without lattice effects, meaning without Umklapp scattering. The first temperature correction to this is of the type $\sqrt{T}$. This has to be contrasted from the standard $1/T^2$ in Fermi liquids (neglecting additional logarithms and above all disorder) or the strictly independent of temperature (up to logarithms) behavior in MLG.
In a second part we studied the Coulomb drag resistivity between two BLGs covering the whole range from deep within the Fermi liquid regime all the way to CN. 
We find that in the Fermi liquid regime drag behaves very similarly to drag in a standard 2DEG or MLG. In contrast to MLG, however, we find saturation of the drag resistance upon decreasing the distance $d$ for physically reasonable parameters.
In the vicinity of the CN point, like in the case of MLG, we find an interesting interplay between interaction effects and disorder making the drag resistivity strongly dependent upon the ratio of the two.

\subsubsection*{Acknowledgements}
We acknowledge collaborations on related problems with M. M\"uller, S. Sachdev, and J. Schmalian. 
This work was supported by the Emmy-Noether program FR 2627/3-1 (LF).

\appendix
\section{The collision integrals}\label{app:collision}
All conductivities are calculated using the variational principle.\cite{zimanbook}
In Ref.~\onlinecite{we} the formalism was adapted to calculate the drag resistivity.
In the following we present the collision integrals resulting from the Hamiltonians shown in section \ref{sec:model}.
They are obtained by the linearization of the Boltzmann equation and taking into account all possible second order 
processes.\cite{Fritz2008}

The collision matrix has three contributions due to the three different scattering mechanisms: intralayer scattering, interlayer scattering, and impurity scattering. 
We label it with the layer index $ a, p $ and particle--hole index $ \l = \pm $, resulting in a $ 4 \times 4 $ matrix.

The layer diagonal part has contributions from all three processes
\begin{equation}
 C^{aa}_{\l \l^\prime} = C_{\l \l^\prime}^{aa, intra} + C_{\l \l^\prime}^{aa, inter} + \delta_{\l \l^\prime} C_{\l \l^\prime}^{aa, imp} \; .
\end{equation}
The contribution from the elastic impurity scattering reads
\begin{widetext}
\begin{align}
C_{\l \l}^{a/p, imp} =  n_{imp} \left( \frac{2 \pi e^2}{\ve T} \right)^2 \intk \intq \frac{\delta(\frac{k^2}{2 m}- \frac{q^2}{2 m})}{(|\bk-\bq| + q_{TF})^2} f^{0, a/p}_\l (k) (1-f^{0, a/p}_\l(q) )
|M_{\l , \l} (\bk, \bq) |^2 \, \bk \cdot \frac{\bk-\bq}{m} \; , 
\end{align}
where $ M $ was defined in Eq.~\eqref{eq:vertex}.
For the intralayer Coulomb scattering we find
\begin{align}
 C_{\l \l^\prime}^{aa, intra}  &  = \frac{2 \pi}{m T^2} \intk \intq \intko \; \delta  \left(  \frac{k^2}{2 m}  -  \frac{k^2_1}{2 m}  - \frac{|\bk+\bq|^2}{2 m} + \frac{|\bko-\bq|^2}{2 m} \right)  R (\bk, \bko, \bq) \nonumber \\
& \times  f^{0, a}_\l (k) f^{0, a}_{\l^\prime} (k_1) (1- f^{0, a}_\l (|\bk+\bq|) )(1-f^{0, a}_{\l^\prime} (|\bko-\bq|) ) \, \bk \cdot \bq \left(  (1-\delta_{\l \l^\prime}) -\delta_{\l \l^\prime} \right) \; ,
\end{align}
where 
\begin{align} \label{eqapp:Rintra}
R &= 4N |T_{+--+}(\mathbf{k},\mathbf{k_{1}},\mathbf{q} )|^{2} +|T_{+-+-}(\mathbf{k},\mathbf{k_{1}},\mathbf{k_{1}}-\mathbf{k}-\mathbf{ q})|^{2} -4T_{+--+}(\mathbf{k},\mathbf{k_{1}},\mathbf{q}) \times \nonumber \\
& \times T_{+-+-}^{\star }(\mathbf{k },\mathbf{k_{1}},\mathbf{k_{1}}-\mathbf{k}-\mathbf{q})  - 
4T_{+-+-}^{\star}(\mathbf{k},\mathbf{k_{1}},\mathbf{k_{1}}-\mathbf{k}-\mathbf{q} )T_{+--+}(\mathbf{k},\mathbf{k_{1}},\mathbf{q}) \; ,
\end{align}
and $ T $ was defined in Eq.~\eqref{eq:Tintra}. The contribution of the interlayer scattering to the layer diagonal part of the collsion matrix reads
\begin{align}
 C_{\l \l^\prime}^{aa, inter}  &  = \frac{2 \pi}{m T^2} \intk \intq \intko \; \Bigg[ \delta  \left(  \frac{k^2}{2 m}  -  \frac{k^2_1}{2 m}  - \frac{|\bk+\bq|^2}{2 m} + \frac{|\bko-\bq|^2}{2 m} \right)  \nonumber \\ 
& \Big[ \, \widetilde{R}_{11} (\bk, \bko, \bq) \times  f^{0, a}_\l (k) f^{0, p}_{\l^\prime} (k_1) (1- f^{0, a}_\l (|\bk+\bq|) )(1-f^{0, p}_{\l^\prime} (|\bko-\bq|) ) \, \bk \cdot \bq \left( -\delta_{\l \l^\prime} \right) \nonumber \\
 & + \widetilde{R}_{12} (\bk, \bko, \bq) f^{0, a}_\l (k) f^{0, p}_{\l^\prime} (k_1) (1- f^{0, p}_\l (|\bk+\bq|) )(1-f^{0, a}_{\l^\prime} (|\bko-\bq|) ) \, \bk \cdot \left( \bk \delta_{\l \l^\prime} - (\bko - \bq) (1-\delta_{\l \l^\prime})\right) \Big] \nonumber \\
 & + \delta  \left(  \frac{k^2}{2 m}  +  \frac{k^2_1}{2 m}  - \frac{|\bk+\bq|^2}{2 m} - \frac{|\bko-\bq|^2}{2 m} \right)   \widetilde{R}_{2} (\bk, \bko, \bq) \, \nonumber \\ 
& \times f^{0, a}_\l (k) f^{0, p}_{\l^\prime} (k_1) (1- f^{0, a}_\l (|\bk+\bq|) )(1-f^{0, p}_{\l^\prime} (|\bko-\bq|) ) \, \bk \cdot \bq \left( -\delta_{\l \l^\prime} \right) \Bigg] \; ,
\end{align}
where
\begin{align} \label{eqapp:Rinter}
\widetilde{R}_{11} & =  4 N |\widetilde{T}_{+--+}(\mathbf{k},\mathbf{k_{1}},\mathbf{q} )|^{2} \; ,\\ 
\widetilde{R}_{12} & =  4 N |\widetilde{T}_{+-+-}(\mathbf{k},\mathbf{k_{1}},\mathbf{k_{1}}-\mathbf{k}-\mathbf{q})|^{2} \; ,\\ 
\widetilde{R}_2    & =  4 N |\widetilde{T}_{++++}(\mathbf{k},\mathbf{k_{1}},\mathbf{q})|^{2} \;, 
\end{align}
and $ \widetilde{T} $ was defined in Eq.~\eqref{eq:Tinter}.  Here only the large N processes contribute\cite{we}.

The layer off--diagonal part has only a single contribution from interlayer scattering, which reads
 \begin{align}
 C_{\l \l^\prime}^{ap}  &  = \frac{2 \pi}{m T^2} \intk \intq \intko \; \Bigg[ \delta  \left(  \frac{k^2}{2 m}  -  \frac{k^2_1}{2 m}  - \frac{|\bk+\bq|^2}{2 m} + \frac{|\bko-\bq|^2}{2 m} \right)  \nonumber \\ 
& \Big[ \, \widetilde{R}_{11} (\bk, \bko, \bq) \times  f^{0, a}_\l (k) f^{0, p}_{\l^\prime} (k_1) (1- f^{0, a}_\l (|\bk+\bq|) )(1-f^{0, p}_{\l^\prime} (|\bko-\bq|) ) \, \bk \cdot \bq \left( 1-\delta_{\l \l^\prime} \right) \nonumber \\
 & + \widetilde{R}_{12} (\bk, \bko, \bq) f^{0, a}_\l (k) f^{0, p}_{\l^\prime} (k_1) (1- f^{0, p}_\l (|\bk+\bq|) )(1-f^{0, a}_{\l^\prime} (|\bko-\bq|) ) \, \bk \cdot \left( \bko (1-\delta_{\l \l^\prime}) - (\bk + \bq) \delta_{\l \l^\prime} \right) \Big] \nonumber \\
 & + \delta  \left(  \frac{k^2}{2 m}  +  \frac{k^2_1}{2 m}  - \frac{|\bk+\bq|^2}{2 m} - \frac{|\bko-\bq|^2}{2 m} \right)   \widetilde{R}_{2} (\bk, \bko, \bq) \, \nonumber \\ 
& \times f^{0, a}_\l (k) f^{0, p}_{\l^\prime} (k_1) (1- f^{0, a}_\l (|\bk+\bq|) )(1-f^{0, p}_{\l^\prime} (|\bko-\bq|) ) \, \bk \cdot \bq \left( \delta_{\l \l^\prime} \right) \Bigg]
\end{align}
\end{widetext}
To calculate the conductivities using the variational principle, we need the driving terms of the relevant modes \cite{zimanbook}.
For the model at hand and the momentum mode, that we have used, they are of the form
\begin{equation} \label{eqapp:driving}
 \bD^a  = (D^a_+, D^a_-) \; , \text{and  } \bD^p = (D^p_+, D^p_-) \; ,
\end{equation}
where
\begin{align} 
D^{a/p}_\pm =  \frac{1}{T} \intk f^{0, a/p}_\pm (1-f^{0, a/p}_\pm) \, \bk \cdot \frac{\bk}{m} \; .
 \end{align}
Then the conductivities can be calculated according to
\begin{eqnarray} 
\s_a & = & \frac{e^2}{h} \, \frac{N \pi}{T} \; (\bD^a , 0 ) \cdot
 \left ( \begin{array} {cc}  C^{aa} & C^{ap} \\ C^{pa}  &  C^{pp}    \end{array} \right)^{-1} \cdot \left( \begin{array}{c}  \bD^a \\ 0  \end{array} \right)  \label{eq:conda_drag} \; , \nonumber \\  
\s_d & = &  \frac{e^2}{h} \, \frac{N \pi}{T} \; (0, \bD^p ) \cdot
 \left ( \begin{array} {cc}  C^{aa} & C^{ap} \\ C^{pa}  &  C^{pp}    \end{array} \right)^{-1} \cdot \left( \begin{array}{c}  \bD^a \\ 0  \end{array} \right)  \label{eq:condd_drag} \; ,\nonumber  \\
\s_p & = &  \frac{e^2}{h} \, \frac{N \pi}{T} \; (0, \bD^p ) \cdot
 \left ( \begin{array} {cc}  C^{aa} & C^{ap} \\ C^{pa}  &  C^{pp}    \end{array} \right)^{-1} \cdot \left( \begin{array}{c}  0 \\\ \bD^p  \end{array} \right)  \; . \label{eq:condp_drag} \nonumber  
\end{eqnarray}
For clarity we have omitted the particle--hole index.

\end{document}